\documentclass[12pt]{article}

\usepackage{putex}
\usepackage{amsmath}
\usepackage{amssymb}
\usepackage{graphicx}
\usepackage{epstopdf}
\usepackage{enumerate}
\usepackage{cite}
\usepackage{tensor}
\usepackage{slashed}
\usepackage{feynmf}
\usepackage{hyperref}
\usepackage{bbold}
\usepackage{dsfont}
\usepackage{mathrsfs}
\usepackage{caption}
\usepackage{subcaption}

\usepackage{youngtab}
\usepackage{amsfonts}
\usepackage{braket}

\newcommand\numberthis{\addtocounter{equation}{1}\tag{\theequation}}
\usepackage[utf8x]{inputenc}
\usepackage{titlesec}
\usepackage{fixltx2e} 
\usepackage{cleveref}
\hypersetup{breaklinks}

\usepackage[usenames,dvipsnames,svgnames,table]{xcolor}

\makeatletter
\renewcommand{\@maketitle}{
\newpage
 \begin{center}%
  {\large\bfseries \@title \par}%
 \end{center}%
 \par} \makeatother

\numberwithin{equation}{section}

\titleformat*{\section}{\large\bfseries}

\begin{document}

\institution{UCLA}{ \quad\quad\quad\quad\quad\quad\quad\quad\quad Mani L. Bhaumik Institute for Theoretical Physics, \cr Department of Physics and Astronomy, University of California, Los Angeles, CA 90095, USA}

\title{Entanglement Entropy with a Time-dependent Hamiltonian}

\authors{Allic Sivaramakrishnan\footnote{\texttt{allic@physics.ucla.edu}}}

\abstract{  The time evolution of entanglement tracks how information propagates in interacting quantum systems. We study entanglement entropy in CFT$_2$ with a time-dependent Hamiltonian. We perturb by operators with time-dependent source functions and use the replica trick to calculate higher order corrections to entanglement entropy. At first order, we compute the correction due to a metric perturbation in AdS$_3$/CFT$_2$ and find agreement on both sides of the duality. Past first order, we find evidence of a universal structure of entanglement propagation to all orders. The central feature is that interactions entangle unentangled excitations. Entanglement propagates according to ``entanglement diagrams,'' proposed structures that are motivated by accessory spacetime diagrams for real-time perturbation theory. To illustrate the mechanisms involved, we compute higher-order corrections to free fermion entanglement entropy. We identify an unentangled operator, one which does not change the entanglement entropy to any order. Then, we introduce an interaction and find it changes entanglement entropy by entangling the unentangled excitations. The entanglement propagates in line with our conjecture. We compute several entanglement diagrams. We provide tools to simplify the computation of loop entanglement diagrams, which probe UV effects in entanglement propagation in CFT and holography.    }

\date{}

\maketitle
\setcounter{tocdepth}{2}
\tableofcontents

\section{Introduction}

Entanglement is a fundamental feature of quantum field theory. The program of studying entanglement and other information-theoretic quantities in field theory has recently led to new understanding of the well-known holographic correspondence between gravitational theories in anti-de Sitter spacetimes (AdS) and large-$N$ strongly-coupled conformal field theories (CFTs). The AdS/CFT correspondence provides a route towards understanding quantum gravitational effects through their dual CFT description, and aspects of strongly coupled field theories through their dual AdS solutions \cite{AdSCFT}. Basic entries in the AdS/CFT dictionary involve information-theoretic quantities, including quantum error correction, complexity, mutual information, and relative entropy \cite{DongHW16,AlmheiriDH14,BrownRSSZ15,HartmanM13, RelativeEntropyEqualsBulkRelativeEntropy}. The black hole information loss paradox is intimately tied to questions of entanglement across the horizon \cite{Firewalls, Harlow14}. 

We will focus on entanglement entropy in this work. A significant body of evidence suggests that surface area in AdS calculates the entanglement entropy of CFT subregions, see for example \cite{RyuTakayanagi, HubenyRangamaniTakayanagi, LewkowyczM13,HRTProof}. CFT entanglement entropy can also be used to derive bulk equations of motion \cite{FaulknerGHMR13}. We investigate the time-dependence of entanglement entropy in CFTs. We work in $1+1$ dimensions, in which global conformal symmetry is enhanced to Virasoro symmetry, providing greater control over the system in question. We study time-dependent perturbations of vacuum entanglement entropy of a single interval in a CFT and apply our results to the AdS$_3$/CFT$_2$ correspondence.

The time evolution of entanglement entropy in excited states has been well-studied, see for example \cite{CasiniHuertaMyers,HeNTW14,NozakiNT14,Caputa14,HartmanJK15, AsplundBGH115,AsplundBGH215,Afkhami-JeddiHKT17}. Excited states can also be created by a time-dependent Hamiltonian, and the two setups are related but distinct \cite{Allic16}. Excited states have been used to model quantum quenches, wherein the Hamiltonian changes abruptly and the state is no longer a vacuum of the new Hamiltonian \cite{Ugajin13,AsplundBGH215}. Studies of entanglement with time-dependent Hamiltonians have also focused on quantum quenches \cite{EislerP07,CalabreseC207,CalabreseC07,CalabreseC05,DasGM14,DasGM15,LokhandeOP17}. General features of Hamiltonian perturbations by local operators have been explored \cite{RosenhausS114,RosenhausS414} and a study of their time-dependence has been initiated through conformal perturbation theory \cite{LeichenauerMS16}. In \cite{LeichenauerMS16}, the first law was used to calculate the change in entanglement entropy to first order in $J$ due to the Hamiltonian perturbation $J(x,t) \mathcal{O}(x)$ for operators $\mathcal{O}$ in the free scalar and fermion theories in various spacetime dimensions. Their CFT results agreed with that of the Hubeny-Rangamani-Takayanagi prescription (HRT) in the bulk, the covariant generalization of the Ryu-Takayanagi prescription (RT) for computing CFT entanglement entropy through extremal surface area.

We will now summarize this manuscript for the reader's convenience. In this work, we study corrections to vacuum entanglement entropy of a single interval $A$ due to Hamiltonian perturbations of the form $J(x,t) \mathcal{O}(x)$. For illustrative purposes, the source function $J(x,t)$ is localized in spacetime, but our methods apply for general source functions. Our primary CFT tool is the replica trick. 

In Section 2, we provide background on the analytic continuation of correlators from Euclidean to Lorentzian signature in position space and the replica trick.

In Section 3, we warm up with first-order Euclidean computations. We consider an infinitesimal Weyl transformation, equivalent to choosing $\mathcal{O}= \text{Tr}(T)$, and compute the change in entanglement entropy with the replica trick and a proper length cutoff procedure. As expected, entanglement entropy changes only at the location of $\mathcal{O}$, the only place conformal symmetry is broken. Using a bulk diffeomorphism that implements the CFT metric transformation, we compute the change in the Ryu-Takayanagi surface area and find agreement with the CFT result. The Euclidean entanglement entropy changes only due to the perturbation at the entangling surface $\partial A$, as in the CFT.

In Section 4, we work in Lorentzian signature and perform a metric perturbation equivalent to choosing $\mathcal{O}(z) = T(z)$. Computing the entanglement entropy in the CFT using the replica trick and the entanglement first law give the same result, and we identify a non-trivial causality property that the modular Hamiltonian obeys but does not make manifest. We compute the change in entanglement entropy in the bulk using the corresponding AdS$_3$ solution and find agreement with the CFT result. The bulk geodesic integral reduces to the CFT first law integral. We conclude that the perturbation changes physics at the entangling surface, just as in the Euclidean case.

Shifting gears, we move to higher orders in perturbation theory. Here, we find evidence of a universal structure of entanglement propagation to all orders: interactions entangle unentangled excitations according to entanglement diagrams. The creation of entanglement through interaction is itself a familiar mechanism in other contexts \cite{CarneyCS16,GrignaniS16,GiddingsR17}; here, we investigate this mechanism in real-time perturbation theory.

\begin{figure}[h]
\begin{center}
\includegraphics[width=0.7 \textwidth]{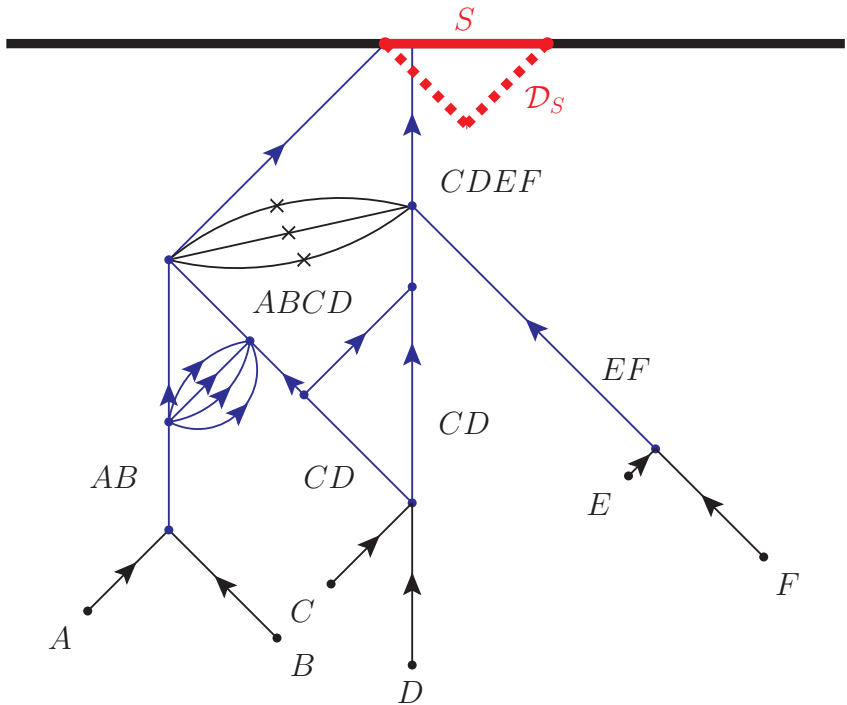}
\end{center}
\caption{\small 
A sample spacetime entanglement diagram for subsystem $S$. $\mathcal{D}_s$ is the domain of dependence of $S$. $A-F$ label incoming unentangled excitations, which interactions subsequently entangle as labeled.
}
\label{EntanglementDiagramIntroduction}
\end{figure}

In Section 5, we make the precise conjecture. For operators $\mathcal{O}$ that do not change entanglement entropy to any order in $J$, certain interactions $\lambda \mathcal{O}_\lambda$ in the Hamiltonian entangle these excitations. Entanglement changes only when a non-trivial ``entanglement diagram'' can be drawn of the process, depicting entanglement propagating through a web of interactions. See figure \ref{EntanglementDiagramIntroduction} for an example. Entanglement diagrams are position-space diagrams associated with the computation of entanglement entropy in real-time perturbation theory, with rules specific to entanglement propagation. However, even in the case of perturbation about a free field theory, entanglement diagrams are not the standard spacetime Feynman diagrams built from Wick contractions of the elementary fields. Instead, lines and vertices in entanglement diagrams are built from operators that serve as building blocks of entanglement in that theory. We provide a procedure to identify these operators.

Entanglement diagrams explicitly differentiate between two mechanisms of entanglement: entanglement due to interactions between excitations, and entanglement due to pre-existing background state correlations. We develop a diagrammatic method of organizing and streamlining real-time perturbation theory computations that makes causality properties manifest.

In Section 6, we perform calculations that provide evidence for the conjecture in Section 5. We use explicit twist operators in the bosonized free fermion theory to compute perturbative corrections to entanglement entropy, as in for example \cite{DattaDFK14}. In the free fermion theory, the natural entanglement diagrams are the position-space Feynman diagrams associated with real-time perturbation theory, but in which lines are Wick contractions of the free boson $\phi$ rather than the free fermion $\psi = e^{i\phi}$. We identify an unentangled operator $\mathcal{O} = \partial \phi$, the spin 1 current, that does not change entanglement entropy to any order. Then, we compute entanglement entropy in the presence of the cubic interaction $\mathcal{O}_\lambda = (\partial \phi)^3$, the spin-three current. Entanglement entropy changes at order $J^2 \lambda^2$, and entanglement propagates according to the associated entanglement diagrams. We compute all $J^2 \lambda^2$ diagrams. The computation manifests various features of the conjecture in Section $6$, for instance that processes that would contribute to a generic correlator are prohibited entanglement entropy corrections according to the entanglement diagram rules. The free fermion is an elementary, tractable testing ground for features of entanglement propagation to high loop order.

Our goal is to make locality and causality manifest and so we express the lightcone divergences in position space rather than momentum space. Divergences can be addressed using methods in \cite{BerensteinM14}. We consider irrelevant, marginal, and relevant $\mathcal{O}_\lambda$, and we expect that the entanglement structure we find is independent of operator dimension. All expectation values are taken in the vacuum unless otherwise specified, and beginning in Section 4, we omit overall numerical factors that play no role in our results.

\section{Background}

\subsection{Analytic continuation to Lorentzian signature}

We review the procedure of analytically continuing correlators from Euclidean to Lorentzian signature, connecting the $i\epsilon$ prescription that is transparent in free field theory to the more general contour prescription discussed in \cite{HartmanJK15}.

We begin with the $i\epsilon$ prescription for obtaining Lorentzian two-point functions in free field theory. We will use the free scalar field in $d+1$ dimensions as an example and use the mostly minus signature. Just as in the Feynman prescription, we may begin with the Euclidean Green's function
\begin{equation}
\mathcal{G}(x) = \int_{\mathcal{C}}d^{d+1} p \frac{e^{-i(p^0 (\pm i \epsilon) -\mathbf{p} \cdot \mathbf{x})}}{(p^0)^2-(\mathbf{p})^2-m^2},
\end{equation}
with contour $\mathcal{C}$ for $p^0$ chosen along the imaginary axis. Notice that if Euclidean time $\epsilon$ is continued to Lorentzian time in the integrand, the integral will diverge unless the contour is rotated. Choosing $-i\epsilon$ allows us to close the contour in the $p^0>0$ half of the plane, enclosing the $E_p=\sqrt{(\mathbf{p})^2+m^2}$ pole. At this point, $\epsilon$ is analytically continued so that $-i \epsilon \rightarrow t - i \epsilon$, which amounts to increasing $t$ to its non-zero Lorentzian value. Taking the $\epsilon \rightarrow 0 $ limit, we obtain the Lorentzian two-point function $\braket{0|\phi(x) \phi(0)|0}$. Using the $i\epsilon$ value instead gives $\braket{0|\phi(0) \phi(x)|0}$. In the spacelike region $x^2 < 0$, choosing either $\pm i \epsilon$ will give the same result. In this case, the operators commute just as in the Euclidean correlator \cite{StreaterWightman,Haag}. 

Now we examine the $i\epsilon$ prescription in position space. Correlators are multivalued in complex time and the $i\epsilon$ prescription indicates the direction from which we approach branch cuts in the correlator when we continue from complex to real time. Consider a spinless operator $\mathcal{O}$ of dimension $\Delta$ in a CFT as an example. In the spacelike region $x^2 < 0$,
\begin{equation}
\braket{0|\mathcal{O}(x) \mathcal{O}(0)|0} = \frac{1}{(-x^2)^{\Delta}}.
\end{equation}
When $-x^2 > 0$, choosing $\pm i\epsilon$ gives the same result because there is no branch cut in this region. When $-x^2$ crosses zero, choosing $\pm i\epsilon$ must give different results. We may orient the branch cut of $(-x^2)^{-\Delta}$ along the negative real axis of $-x^2$. To obtain the correlator for timelike separation, we must analytically continue the spacelike expression.
\begin{equation}
\braket{\mathcal{O}(\mathbf{x},t\pm i\epsilon) \mathcal{O}(0)} = \frac{1}{((|\mathbf{x}|+t\pm i\epsilon)(|\mathbf{x}|-t\mp i\epsilon))^{\Delta}} 
\approx
\frac{1}{(\mathbf{x}^2-t^2 \mp i\epsilon ~\text{sgn}(t))^{\Delta}}.
\end{equation}
The quantity $\braket{\mathcal{O}(\mathbf{x},t\pm i\epsilon) \mathcal{O}(0)}$ denotes the continuation of the Euclidean two-point function to complex time, and so the operator ordering of this expression is meaningless except in the limit $\epsilon \rightarrow 0$. Moving $\mathcal{O}(x)$ through the future $f$ or past $p$ lightcone of $\mathcal{O}(0)$ corresponds to $t$ crossing $\pm |\mathbf{x}|$, which is equivalently fixed by $\text{sgn}(t)$. This choice is separate from the choice of $\pm i\epsilon$ that gives the two different operator orderings. The timelike two-point function in the two kinematic regions $t>|\mathbf{x}|$ and $t<-|\mathbf{x}|$ is
\begin{align*}
\braket{\mathcal{O}(\mathbf{x},t\pm i\epsilon) \mathcal{O}(0)}_f &=
\frac{1}{(t^2-\mathbf{x}^2)^{\Delta}} e^{\pm \pi \Delta i },
\\
\braket{\mathcal{O}(\mathbf{x},t\pm i\epsilon) \mathcal{O}(0)}_p &=
\frac{1}{(t^2-\mathbf{x}^2)^{\Delta}} e^{\mp \pi \Delta i}.
\numberthis
\label{TimelikeTwoPointPrescription}
\end{align*}
The $i\epsilon$ prescription is precisely what chooses the branch cut of the two-point function when the argument becomes negative. There is an unphysical choice of overall normalization. Here we have chosen $e^{\pm i \pi}$ to parameterize the two directions of approaching the branch cut, but in general we may choose $e^{\alpha i},e^{\beta i}$ for any $|\alpha-\beta| = 2 \pi$. A common choice is $\alpha=0, \beta= 2\pi$. 

Having addressed the two-point function, we summarize the rules for obtaining the $n$-point correlator in which all operators are timelike separated from one another. The Lorentzian correlator
\begin{equation}
\braket{\mathcal{O}_1(x_1) \mathcal{O}_2(x_2) \mathcal{O}_3(x_3) \ldots}
\end{equation}
corresponds to the continuation $t_i \rightarrow t_i -i\epsilon_i$ with $\epsilon_1>\epsilon_2>\epsilon_3\ldots > 0 $ of the Euclidean correlator in the limit of $\epsilon_1 \rightarrow 0$  \cite{StreaterWightman, Haag, LuscherM75, OsterwalderS73, OsterwalderS75}. Moving $\mathcal{O}_i(x_i)$ past $\mathcal{O}_j(x_j)$ in the correlator amounts to reversing the sign of $\epsilon_i-\epsilon_j$, which as we have seen amounts to approaching the branch cut in $x_i-x_j$ from the opposite direction, as in \eqref{TimelikeTwoPointPrescription}. It is equivalent to view the $i\epsilon$ prescription of approaching branch cuts as the complex time path of analytic continuation passing through different sheets of the correlator before reaching the real axis \cite{HartmanJK15}. The location of the singularities can be different on different sheets of the correlator. See \cite{HartmanJK15} for a detailed explanation of this contour prescription.

Now, we address the null singularities of correlators. Correlators can have delta-function lightcone singularities that require a generalization of the $i\epsilon$ prescription. We review the lightcone singularity of the $3+1$ dimensional free scalar two-point function, as in for example \cite{GreinerQED}, and then provide its generalization to all dimensions. To our knowledge this generalization has not been stated in position space in the literature, so we justify the result in detail.

We begin by calculating the spacelike two-point function $D(x)$ in $d+1$ dimensions. We will omit most numerical prefactors. Performing the $p^0$ integral,
\begin{equation}
D(x) =\int d^d p \frac{1}{E_p} e^{\mp i E_p t+ i \mathbf{p} \cdot \mathbf{x}}, ~~~~~~ E_p = |\mathbf{p}|.
\end{equation}
The $\pm$ signs specify the pole $\pm E_p$ enclosed. With $r \equiv  |\mathbf{p}|$, we can integrate over the angular directions. The lightcone divergence comes from the large $r$ behavior of $D(x)$, which is
\begin{equation}
D(x) = \frac{1}{|\mathbf{x}|}\int_{0}^{\infty} dr r^{d-3}
\left(
e^{ -i r (\pm t-|\mathbf{x}|)} - e^{ -i r (\pm t+|\mathbf{x}|)}
\right).
\label{GreinerNullSingularity}
\end{equation}
In the familiar case of $d=3$ \cite{GreinerQED,BirrellDavies}, the quantity \eqref{GreinerNullSingularity} is
\begin{equation}
D(x) = \frac{1}{|\mathbf{x}|}(\delta(\pm t-|\mathbf{x}|)-\delta( \pm t+|\mathbf{x}|)) = \delta(t^2-|\mathbf{x}|^2) = \delta(x^2).
\end{equation}
We could have obtained this divergence by analytically continuing the Euclidean two-point function instead. For $t>0$,
\begin{equation}
\lim_{\epsilon \rightarrow 0} D(\mathbf{x},t \pm i\epsilon) = \lim_{\epsilon \rightarrow 0} \frac{1}{x^2 \mp i\epsilon} = \pm i \pi \delta(x^2) + \text{p.v.}\left( \frac{1}{-x^2} \right), 
\end{equation}
where p.v. denotes the Cauchy principal value. We will omit the principal value designation from now on, but it is understood to be present and can be restored easily. For higher dimensions, \eqref{GreinerNullSingularity} becomes
\begin{equation}
D(x) = \frac{1}{|\mathbf{x}|} \partial_{t}^{d-3} \int dr 
\left(
e^{ -i r(\pm t-|\mathbf{x}|)} - e^{ -i r (\pm t+|\mathbf{x}|)}
\right)
=
\partial_{t}^{d-3} \delta(x^2).
\label{4DNullSingularity}
\end{equation}
Using 
\begin{equation}
\lim_{\epsilon \rightarrow 0} \frac{1}{(s \pm i\epsilon)^{n}}=\frac{(-1)^{n-1}}{(n-1)!}\partial_{s}^{n-1} \lim_{\epsilon \rightarrow 0} \frac{1}{s \pm i\epsilon},
\end{equation}
and \eqref{TimelikeTwoPointPrescription}, we can write the full CFT two-point function of $\mathcal{O}$ in general dimensions by analytically continuing the position-space Euclidean correlator.
\begin{align*}
\text{Spacelike}~~~ x^2 < 0: ~~~&\braket{\mathcal{O}(x) \mathcal{O}(0)} =\braket{\mathcal{O}(0)\mathcal{O}(x) } = \frac{1}{(-x^2)^{\Delta}}.
\\
\text{Timelike, null} ~~~ x^2 \geq 0: ~~~&
\braket{\mathcal{O}(x) \mathcal{O}(0)} = 
\braket{\mathcal{O}(\mathbf{x},t - i\epsilon) \mathcal{O}(0)},
\\
&\braket{\mathcal{O}(0) \mathcal{O}(x)} = 
\braket{\mathcal{O}(\mathbf{x},t + i\epsilon) \mathcal{O}(0)},
\numberthis
\end{align*}
where for $x^2 \geq 0$,
\begin{align*}
\braket{\mathcal{O}(\mathbf{x},t\pm i\epsilon) \mathcal{O}(0)}_f &=
 \pm \frac{(-1)^{\Delta-1}}{(\Delta-1)!} \cdot i \pi \partial_{x_f}^{\Delta-1} \delta(x^2)
+
\frac{1}{(x^2)^{\Delta}} e^{\pm \pi \Delta i },
\\
\braket{\mathcal{O}(\mathbf{x},t\pm i\epsilon) \mathcal{O}(0)}_p &=
 \mp \frac{(-1)^{\Delta-1}}{(\Delta-1)!} \cdot i \pi \partial_{x_p}^{\Delta-1} \delta(x^2)
+
\frac{1}{(x^2)^{\Delta}} e^{\mp \pi \Delta i}.
\numberthis
\label{TwoPointPrescription}
\end{align*}
The lightcone coordinates $x_f = |\mathbf{x}|-t$ and $x_p = |\mathbf{x}|+t$. The prescription \eqref{TwoPointPrescription} matches the $3+1$ dimensional free-scalar result of \cite{BirrellDavies}.

We have discussed correlators of bosonic operators, but there is an additional negative sign involved in computing fermionic correlators through analytic continuation. Taking $t_z \rightarrow t_z-i\epsilon$ in the two-point function $\braket{0|\psi(z) \psi^*(w)|0}$ at timelike separation produces $1/(z-w)$, but to obtain the other ordering one must take $t_z \rightarrow t_z +i\epsilon$ and also anticommute the fermions, obtaining $\braket{0|\psi^*(w) \psi(z)|0} = -1/(z-w)$. In even spacetime dimensions, the commutator of free scalars and anticommutator of free fermions have support only on the lightcone while the anticommutator of scalars and commutator of scalars have support everywhere inside the lightcone. However, this is reversed in odd spacetime dimensions, leading to what is known as a violation of Huygen's principle in odd dimensions \cite{Ooguri86}.

\subsection{Entanglement entropy from the replica trick}

Our main tool in this work will be the replica trick, a useful method of calculating entanglement entropy in $1+1$ dimensional CFTs. We will briefly review the replica trick here, but for a more complete review, see for example \cite{ReplicaTrick, CalabreseC09}. The entanglement entropy $S_A$ of subsystem $A$ can be computed as $S_A = -\partial_{n=1} \text{tr}(\rho_A)^n,$ where $\rho_A$ is the reduced density matrix of $A$. The quantity $\text{tr} \rho_A^n$ can be computed from an $n$-sheeted Riemann surface,
\begin{equation}
\text{tr} \rho_A^n = \frac{Z_n(A)}{(Z_1)^n},
\label{RenyiEntropy}
\end{equation}
where $Z_n(A)$ is the partition function of the theory on an $n$-sheeted Riemann surface. The surface is formed by joining $n$ copies of the plane along cuts located at $A$ on each sheet. The theory on the Riemann surface can be mapped to $n$ copies of the theory on the plane with local twist operators that impose the correct boundary conditions upon the $n$ species of replica fields. $\text{tr}\rho_A^n$ is proportional to the expectation value of these twist operators \cite{ReplicaTrick}. The twist operator $\Phi_n(u)$ and anti-twist operator $\bar{\Phi}_n(v)$ are inserted when the subsystem $A$ is the interval $(u,v)$, the case we consider. We use the twist operator normalization fixed by $\braket{\Phi_n(u) \bar{\Phi}_n(v)} = 1/(u-v)^{2\Delta_{\Phi}}$, where the scaling dimension $\Delta_\Phi = \frac{c}{12}(n-\frac{1}{n})$ and $c$ is the central charge of the theory.

The action of the twist operators is apparent in their diagonalizing basis. Labeling the $n$ replica fields on the plane as $\phi_l$, the diagonalizing replica fields $\tilde{\phi}_k$ are
\begin{equation}
\tilde{\phi}_k = \sum_{l=0}^{n-1} e^{2 \pi i l \frac{k}{n}} \phi_l,
~~~~~
k = 0,1,\ldots,n-1.
\numberthis
\end{equation}
Moving the $\phi_l$ around the twist operator $\Phi_n(u)$ takes $\phi_l \rightarrow \phi_{l\pm 1}$, and is equivalent to multiplying $\tilde{\phi}_k$ by $e^{2\pi i k/n}$. Moving around the anti-twist $\bar{\Phi}_n(v)$ produces a factor of $e^{-2\pi i k/n}$. Deforming the Lagrangian $\mathcal{L}$ of the original theory by $\phi^m$ corresponds to deforming the replicated Lagrangian $\mathcal{L}_n$ by $\sum_l (\phi_l)^m$, and
\begin{equation}
\sum_{l=1}^n (\phi_l)^m = \delta_{0,\sum k_i} \tilde{\phi}_{k_1} \tilde{\phi}_{k_2} \ldots \tilde{\phi}_{k_m},
\label{MultilinearOperatorReplicaBasisConversion}
\end{equation}
up to an overall $n$-dependent normalization factor. For bilinears, $\sum_l \phi_l \phi_l = \sum \tilde{\phi}_k \tilde{\phi}_{-k}$.

The free fermion theory provides an explicit realization of the twist operators through bosonization. Details of the setup not provided here can be found in \cite{DattaDFK14, CasiniFH05, CasiniH09, CasiniH08}. In bosonization, the holomorphic and anti-holomorphic fermions are written in terms of a free scalar as follows:
\begin{equation}
\psi(z) = e^{i \phi(z)},
~~~~~~~~~
\bar{\psi}(\bar{z}) = e^{i \bar{\phi}(\bar{z})}, 
\end{equation}
where $\braket{\phi(z) \phi(w)} = -\ln(z-w)$ and similarly for $\bar{\phi}$. Here $\phi, \bar{\phi}$ are real.

Under cyclic permutation of $\psi_l$, the fermion on the last sheet must be identified with the first fermion up to a negative sign that depends on whether $n$ is even or odd, as $\psi_l$ can always be redefined to eliminate all but a possible overall sign change under this operation \cite{CasiniH09}. The diagonalizing replica fields $\tilde{\psi_k}$ are
\begin{equation}
\tilde{\psi}_k = \sum_{l=0}^{n-1} e^{2 \pi i l \frac{k}{n}} \psi_l,~~~~~
k = -1/2(n-1),-1/2(n-1)+1,\ldots,1/2(n-1).
\label{DiagonalizingReplicaFields}
\end{equation}
The inverse transformation is
\begin{equation}
\psi_l = \frac{1}{n}\sum_{k} e^{-i 2 \pi l \frac{k}{n}}\tilde{\psi}_{k}.
\label{ReplicaFields}
\end{equation}
We will be using fermion bilinears in this work, for which $\sum_l \psi_l \psi_l^* = \sum \tilde{\psi}_k \tilde{\psi}_{k}^*$. The twist operators for the free fermion are 
\begin{align*}
\Phi_n(z, \bar{z}) &= \prod_k e^{i \frac{k}{n} (\phi_{k}(z) - \bar{\phi}_{k}(\bar{z}) )}, 
\\
\bar{\Phi}_n(z, \bar{z}) &= \prod_k  e^{-i \frac{k}{n} (\phi_{k}(z) - \bar{\phi}_{k}(\bar{z}) )} .
\numberthis
\end{align*}
The operator $e^{i \alpha \phi(z)}$ has conformal dimension $\frac{1}{2}\alpha^2$, consistent with the normalization of $\braket{\phi(z) \phi(w)}$.

We will use the normal-ordered product of twist operators. Notice that this operator is a neutral product of vertex operators. For interval $A$ between $(u,\bar{u})$ and $(v,\bar{v})$, 
\begin{equation}
N(\Phi_n(u,\bar{u}) \bar{\Phi}_n(v,\bar{v}))
=
:\text{exp}
\left(
\sum_k
i\frac{k}{n}
\left[
\phi_k(u)-\bar{\phi}_k(\bar{u})
-
\phi_k(v)+\bar{\phi}_k(\bar{v})
\right]
\right)
Z_0(n),
\label{NormalOrderedTwists}
\end{equation}
where $Z_0(n)$ is the $n$-sheeted partition function for the vacuum \cite{CasiniH09}. $Z_0(1) = 1$ and $-\partial_n|_{n=1} Z_0(n) = S_A$, the entanglement entropy of the interval in the vacuum. The factor $Z_0(n)$ will not contribute to any of our entanglement entropy calculations, so we will omit it. We will sometimes use the notation $\Phi_n(u, \bar{u}) = \Phi_n(u)$ for compactness.

\section{First order metric perturbation: Euclidean $AdS_3/CFT_2$}

In this section we work in Euclidean $AdS_3/CFT_2$ and compute the perturbative correction to the vacuum entanglement entropy, which is
\begin{equation}
S_A = \frac{c}{3} \ln\left( \frac{u-v}{\epsilon} \right),
\end{equation}
to first order in a metric perturbation. Here, $\epsilon$ is a UV cutoff. We consider an infinitesimal Weyl transformation as our metric perturbation:
\begin{equation}
 g_{\mu \nu}' = \delta_{\mu \nu} + \omega(x) \delta_{\mu \nu}, ~~~~~ \omega(x) \ll 1.
\end{equation}
We compute the change in entanglement entropy using the replica trick in the CFT and find that the correction depends only on the metric perturbation at the interval's endpoints, which is expected as perturbation by the trace of the stress tensor preserves conformal symmetry wherever $\omega(x) = 0$. Next, we compute the change in entanglement entropy as the change in a proper-length cutoff and find agreement with the replica trick result. Using the Ryu-Takayanagi prescription, we compute the first-order correction in AdS$_3$ and reproduce the CFT result. Our ultimate goal is to work in Lorentzian signature, and this Euclidean computation will mirror features of later Lorentzian calculations.

\subsection{CFT$_2$: Replica trick}
We begin by using the replica trick. The quantity $\text{tr} \rho_A^n$ changes under an infinitesimal Weyl transformation as
\begin{equation}
\delta \text{tr} \rho_A^n
=
-\frac{1}{2} \int d^2 x 
\frac{
\braket{\Phi_n(u) \bar{\Phi}_{n}(v) T^\mu_\mu (x)}\omega(x)
}
{\braket{\Phi_n(u) \bar{\Phi}_{n}(v)}}.
\end{equation}
We have used the stress tensor normalization such that for an infinitesimal diffeomorphism that acts as $g'_{\mu \nu} = g_{\mu \nu} + \delta g_{\mu \nu}$, the action changes as $\delta S = - \frac{1}{2} \int T^{\mu \nu}\delta g_{\mu \nu}$. Using the Ward identity,
\begin{equation}
T^\mu_\mu(x) \Phi_{n}(u) \bar{\Phi}_n(v) = 
-\delta(x-u) \Delta_\Phi  \Phi_n(u) \bar{\Phi}_{n}(v)
-\delta(x-v) \Delta_\Phi \Phi_n(u) \bar{\Phi}_{n}(v) .
\end{equation}
We therefore have
\begin{equation}
\delta \text{tr} \rho^n_A = \frac{c}{24}\left( n-\frac{1}{n}\right)(\omega(u)+\omega(v)),
\end{equation}
and the change in entanglement entropy
\begin{equation}
\delta S_A = \frac{c}{12}(\omega(u) + \omega(v)).
\label{EuclideanTwistEE}
\end{equation}
Only the Weyl transformation at the endpoints changes the entanglement entropy to this order.

\subsection{CFT$_2$: Proper length cutoff}

We can view the change in entanglement entropy as a change in the coordinate length UV cutoff $\epsilon$ defined by a proper length $\epsilon_p$ that is held fixed under the infinitesimal Weyl transformation. After the Weyl transformation, the coordinate length cutoff $\epsilon$ associated with $u$ for example is  
\begin{equation}
\epsilon = \int^{u \pm \epsilon_p}_u ds'
=
\int^{u \pm \epsilon_p}_u \sqrt{e^{\omega} dx^2} 
 \approx \int^{u \pm \epsilon_p}_u \left( 1+\frac{1}{2} \omega(x) \right) dx.
\end{equation}
Assuming $ (\epsilon_p)^n \frac{d^n}{dx^n} \omega(x)  \ll 1$ for $n \geq 1$, we can expand $\omega(x)$ about $u$ and neglect all but the leading term $\omega(u)$. Therefore,
\begin{equation}
\epsilon= \left(1+\frac{1}{2}\omega(u) \right) \epsilon_p.
\label{CutoffTransformation}
\end{equation}
We remind the reader that we have the freedom to choose two distinct UV cutoffs, one associated with each endpoint \cite{Holzhey94}. Choosing two cutoffs $\epsilon_p(u), \epsilon_p(v)$ is equivalent to replacing $\epsilon_p$ by $\sqrt{\epsilon_p(u) \epsilon_p(v)}$ in the vacuum result with a single cutoff.
\begin{equation}
S_A = \frac{c}{6}
\left(
\ln\left(\frac{u-v}{\epsilon_p(u)}\right)+\ln\left(\frac{u-v}{\epsilon_p(v)}\right)
\right).
\end{equation}
Using \eqref{CutoffTransformation},
\begin{equation}
S_A + \delta S_A \approx \frac{c}{3}
\ln\left(\frac{u-v}{\epsilon}\right) + \frac{c}{12}(\omega(u) + \omega(v)),
\label{EuclideanCutoffEE}
\end{equation}
in agreement with \eqref{EuclideanTwistEE}.

\subsection{AdS$_3$: Ryu-Takayanagi}

For a holographic CFT$_2$, the Ryu-Takayanagi prescription can be used to calculate the change in entanglement entropy in the dual AdS$_3$. In AdS$_3$, we can implement the infinitesimal boundary Weyl transformation through a bulk diffeomorphism. We work in the Poincare patch of AdS$_3$ and use the Fefferman-Graham coordinates for the metric near the boundary.
\begin{equation}
ds^2=d\eta^2+e^{2\eta/l} dz d\bar{z},
\end{equation}
where $\eta \rightarrow \infty$ corresponds to the boundary of AdS$_3$ and $l$ is the AdS radius. In an asymptotically AdS$_3$ spacetime, the CFT$_2$ metric and expectation value of the CFT$_2$ stress tensor can be read off from the form
\begin{equation}
ds^2=d\eta^2+e^{2\eta/l}g_{ij}^{(0)} dz^i dz^j + g_{ij}^{(2)}dx^i dx^j.
\end{equation}
The term $g_{ij}^{(0)}$ is the CFT$_2$ metric and $g_{ij}^{(2)}$ is proportional to the expectation value of the boundary stress tensor \cite{BalasubramanianK99}.

Precisely speaking, conformal transformations of the CFT are a conformal coordinate transformation followed by a Weyl transformation to remove the conformal factor. By modifying the bulk diffeomorphism that produces this boundary conformal transformation \cite{BalasubramanianK99}, we can find the diffeomorphism that will implement only the boundary Weyl transformation. We work in (anti) holomorphic coordinates $z,\bar{z}$ with $z=x+i \tau, \bar{z}=x-i \tau$.  Consider the infinitesimal Weyl parameter
\begin{equation}
\omega = (\epsilon+\bar{\epsilon})/l,
\end{equation}
where $\omega \ll 1$. $\epsilon/l$ will be the small parameter in the bulk. The diffeomorphism that produces the infinitesimal boundary Weyl transformation is
\begin{equation}
z\rightarrow z + \frac{1}{2} l \bar{\epsilon}' e^{-2 \eta/l},
~~~~~
\bar{z} \rightarrow \bar{z} + \frac{1}{2} l \epsilon' e^{-2 \eta/l},
~~~~~
\eta\rightarrow \eta+ \frac{1}{2}(\epsilon+\bar{\epsilon}),
\label{WeylDiffeomorphism}
\end{equation}
where the primes denote (anti)holomorphic derivatives. To first order in $\epsilon, \bar{\epsilon}$,
\begin{equation}
ds^2=d\eta^2+e^{2\eta/l}\left(1+(\epsilon + \bar{\epsilon})/l \right)dz d\bar{z} +\frac{1}{2}l(\epsilon''+\bar{\epsilon}'')dz d\bar{z}.
\end{equation}
It will be convenient to work in Poincare coordinates, with the radial coordinate $\rho = l e^{-\eta/l}$. The transformation \eqref{WeylDiffeomorphism} in Poincare coordinates is
\begin{equation}
z \rightarrow z +\frac{\rho^2}{2l}\bar{\epsilon}',
~~~~~
\bar{z} \rightarrow \bar{z} +\frac{\rho^2}{2l}\epsilon',
~~~~~
\rho \rightarrow \rho\left(1-\frac{\epsilon+\bar{\epsilon}}{2l}\right).
\end{equation}
Converting back to coordinates $x,\tau$, the transformation \eqref{WeylDiffeomorphism} is
\begin{equation}
x\rightarrow x + \frac{\rho^2}{4l}(\bar{\epsilon}'+\epsilon'),
~~~
\tau \rightarrow \tau + \frac{\rho^2}{4il}(\bar{\epsilon}'-\epsilon'),
~~~
\rho \rightarrow \rho\left(1-\frac{\epsilon+\bar{\epsilon}}{2l}\right).
\label{PoincareWeylTransformation}
\end{equation}
According to the Ryu-Takayanagi prescription, the geodesic distance between boundary points $x=u, x=v$ computes entanglement entropy. As \eqref{PoincareWeylTransformation} is simply a diffeomorphism, the geodesic distance between two points on the boundary will take the same form before and after the diffeomorphism. The change in length will arise from applying the diffeomorphism to the geodesic length expression.

We will review the geodesic length computation. Consider a $\tau=0$ geodesic without loss of generality. The boundary cutoff surface is at $\rho = \delta \ll l,r$, where $\delta$ is related to the CFT cutoff and $r=v-u$. Geodesics in Euclidean AdS$_3$ are semi-circles. We parametrize the semi-circle centered at $x = a,\rho=0$ and with radius $r$ as
\begin{equation}
\rho=r \sin \theta, ~~~~~~~x=a+r \cos \theta.
\end{equation}
Integrating from $x = a+r$ to $x = a-r$, the geodesic length $L$ is
\begin{equation}
L = 
\int{ds} 
= \int_0^{\pi} d\theta \frac{l}{\sin \theta}  
= l \ln \left( \tan \left(\frac{\theta}{2}\right)\right) \bigg|^{\pi-\delta/r}_{\delta/r}
\approx l \ln (2 r/ \delta)+l \ln (2 r/ \delta).
\end{equation}
The regulated endpoints of the interval, $(x,\rho) = (a\pm r, \delta)$, transform to first order in the $\delta,\epsilon$ as follows, where we denote spatial dependence of $\epsilon, \bar{\epsilon}$ with $[\ldots]$:
\begin{equation}
 (a \pm r,\delta) \rightarrow \left(a \pm r,\delta \left(1-\frac{\epsilon[a \pm r] + \bar{\epsilon}[a \pm r]}{2 l}\right)\right).
\end{equation}
The geodesic length is therefore
\begin{equation}
L= l \ln\left(\frac{2r}{\delta \left(1-\frac{\epsilon[a-r] + \bar{\epsilon}[a-r]}{2 l}\right)}\right)
+l \ln\left(\frac{2r}{\delta \left(1-\frac{\epsilon[a+r] + \bar{\epsilon}[a+r]}{2 l}\right)}\right).
\label{EuclideanGeodesicLength}
\end{equation}
This is the transformed geodesic length. We now use the Ryu-Takayanagi prescription $S_A = \text{Area}_{min}/4G_N$ and the Brown-Henneaux relation $c=\frac{3l}{2G_N}$ \cite{RyuTakayanagi,BrownHenneaux} and find
\begin{equation}
\delta S_A = \frac{c}{12}(\omega(u)+\omega(v)),
\label{EuclideanAdSEE}
\end{equation}
in agreement with the CFT$_2$ result \eqref{EuclideanTwistEE}. The bulk computation resembles the proper-length CFT computation in that the transformation of the cutoff led to the change in entanglement entropy \eqref{EuclideanCutoffEE}. There is no fundamental obstacle to extending \eqref{EuclideanAdSEE} to higher orders in $\omega$: the diffeomorphism that produces the finite Weyl transformation of the boundary is known \cite{Krasnov01}, and correlators of twist operators with stress tensor insertions are fixed by Ward identities.

\section{First order metric perturbation: Lorentzian $AdS_3/CFT_2$}

In this section, we will calculate the first-order change in entanglement entropy due to a metric perturbation. From this point on, we will work entirely in Lorentzian signature. We will view the corresponding Hamiltonian perturbation by the stress tensor as creating an excited state. The two descriptions are entirely equivalent \cite{Allic16}. We consider the excited state $U\ket{0}$ with
\begin{equation}
U = \mathcal{T} \left( e^{-i \int{d^2y g(y^\mu) T(y_-)}} \right),
\label{MetricExcitationUnitary}
\end{equation}
where $y_\pm$ are lightcone coordinates. Our notation in this section is $y_-=y-t_y$ with $y^\mu = (t_y,y)$. The function $g(y^\mu) \equiv g(t_y,y)$ is bounded and contains a small dimensionless parameter so that the first order correction in $g$ to correlation functions dominates for small value of this parameter. To first order in $g$, time ordering will not be relevant. Acting with the operator in \eqref{MetricExcitationUnitary} is equivalent to perturbing the metric by $\delta g_{--} = g(y^\mu)$. We compute entanglement entropy of a constant-time interval $A$ at a time $t_x$ when the source has turned off: $g(t_x,y) = 0$ for all $y$. At this point the Hamiltonian is once again equal to the unperturbed Hamiltonian, but the state is no longer the vacuum of that Hamiltonian. We choose $t_x=0$ for convenience.

The first order correction to entanglement entropy is found by calculating the entanglement entropy in the state $(1-i \lambda T(y_-)) \ket{0}$ to first order in $\lambda$ and then integrating this quantity $\mathcal{I}_\lambda $ against $g$:
\begin{equation}
\delta S_A = \int d^2 y g(y^\mu) \mathcal{I}_\lambda(y^\mu).
\end{equation}
Agreement between CFT and bulk methods occurs for the kernel of the perturbation $\mathcal{I}_\lambda $ as expected. Non-analyticity of $g$ in time poses no fundamental obstacle to defining or implementing the replica trick, and our calculation demonstrates this perturbatively. In the CFT, we will use the entanglement first law and the replica trick and find agreement. A basic causality property of entanglement entropy, that excitations localized to the causal domain of $A$ cannot change the entanglement entropy $S_A$, is not manifest in the modular Hamiltonian. However, this property holds nevertheless, and imposes constraints on the modular Hamiltonian. In the bulk we will use the Hubeny-Rangamani-Takayanagi proposal and reproduce the CFT result. Finally, we will integrate $\mathcal{I}_\lambda $ and provide an interpretation for the change in entanglement entropy as a changing of the physics at the cutoff (entangling) surface. 

\subsection{CFT$_2$: Entanglement First Law}

We will compute the correction to the entanglement entropy using the entanglement first law \cite{CasiniHuertaMyers}. The first-order correction $\delta S_A$ is
\begin{equation}
\delta S_A = \int d^2 y g(y^\mu) \int_A d x f(x) \braket{0|[T_{--}(y_-),T_{00}(x)]|0} 
\label{FirstLawBeginning}
\end{equation}
For the interval $A$, $f(x) = (x-u)(x-v)/(u-v)$. The commutator $\braket{0|[T_{--}(y_-),T_{00}(x)]|0}$ is fixed by conformal invariance.
\begin{equation}
\braket{T_{--}(x,t) T_{00}(0)} = \frac{1}{(x-t)^4},
\end{equation}
where we have omitted the overall factor of $c$. The commutator has support only on the lightcone, as seen from \eqref{TwoPointPrescription}.
\begin{equation}
\braket{[T_{--}(x,t),T_{00}(0)]} = \partial_x^3\delta(x-t).
\end{equation}
The correction to entanglement entropy \eqref{FirstLawBeginning} is
\begin{equation}
\mathcal{I}_\lambda =
\int_{\partial A} dx f(x)
\partial_x^3 \delta(x_--y_-)
\end{equation}
From this expression, it naively seems that entanglement entropy can change when the perturbation is within the causal domain $\mathcal{D}_A$ of $A$, violating a well-known causality property of entanglement entropy \cite{Headrick14, Allic16}. $\mathcal{D}_A$ is defined as the region through which no timelike geodesic can intersect without also passing through $A$. In the present case, $\mathcal{D}_A$ is the causal diamond of $A$. Integrating by parts,
\begin{equation}
\mathcal{I}_\lambda=
-f'(x)
\partial_x \delta(x_--y_-)\bigg|_{\partial A}
+
f''(x)
\delta(x_--y_-)\bigg|_{\partial A}
,
\end{equation}
where primes denote spatial derivatives. We have used that $f(x)$ is zero for $x \in \partial A$, the boundary of the interval, which is necessary for the modular flow to vanish at $\partial A$. If we restored the overall numerical factors we have omitted, we would find that the $i$ in the commutator multiplies the $i$ coming from the perturbation to give a real result.

We have also used that $f'''(x) = 0$ for $x \in A$. We had no reason \textit{a priori} to require $f'''(x) = 0$, but notice that if this were not true, entanglement entropy would change due to an excitation localized entirely within $\mathcal{D}_A$. We see that simple causality considerations restrict the form of $f(x)$. This argument applies only in two dimensions, but the same $f(x)$ appears in higher-dimensional modular Hamiltonians, and so can be viewed as a constraint on the modular Hamiltonian. This argument applies whenever the modular Hamiltonian is given by an integral over the stress tensor over any spacelike surface with the same boundary as $A$. Showing that $f'''(x) = 0$ for $x \in A$ follows from some basic principle would be illuminating. In general, this property holds for the vacuum modular Hamiltonian defined by choosing any Cauchy surface for $\mathcal{D}_A$. Substituting for $f(x)$, we have
\begin{equation}
\mathcal{I}_\lambda =
-\partial_y (\delta(v-y_-)+\delta(u-y_-))
+
\frac{2}{u-v} (\delta(v-y_-)-\delta(u-y_-)).
\label{FirstLawEE}
\end{equation}
While the perturbation we have shown is right-moving, a general metric perturbation can change the entanglement entropy when null-separated from the interval's endpoints.

\subsection{CFT$_2$: Replica Trick}
We compute $\mathcal{I}_\lambda$ using the replica trick \cite{ReplicaTrick} in Lorentzian signature. In Euclidean signature, the expectation value of $T(y_-)$ on the $n$-sheeted Riemann surface is
\begin{equation}
\braket{T(y_-)}_{\mathcal{R}_n} = \frac{\braket{T(y_-) \Phi_n(u) \bar{\Phi}_{n}(v)}}
{\braket{\Phi_n(u) \bar{\Phi}_{n}(v)}}
=
\frac{c}{24}(n-1/n) \frac{(u-v)^2}{(y_--u)^2 (y_--v)^2}.
\end{equation}
Acting with $-\partial_{n=1}$,
\begin{equation}
\partial_n|_{n\rightarrow 1} \braket{T(y_-)}_{\mathcal{R}_n} = \frac{c}{12}
\frac{(u-v)^2}{(y_--u)^2 (y_--v)^2}
\end{equation}

We will assume that we can use the replica trick to calculate the change in entanglement entropy by treating the twist operators as well-defined local operators purely in Lorentzian signature. Using standard real-time perturbation theory implies that corrections to their expectation value will involve computing their commutators with Hamiltonian perturbations. This assumption is the natural sibling of the assumption made in order to compute excited-state entanglement entropy using twist operator insertions \cite{AsplundBGH215}.

Our assumption should not be confused with assuming a purely-Lorentzian definition of the twist operators. The twist operators are ordinarily defined by imposing boundary conditions that lead to a Euclidean $n$-sheeted Riemann surface - moving a diagonalizing replica field $\phi_n$ on sheet $n$ around the twist operator in Euclidean signature exchanges the field for one on another sheet: $\phi_n \rightarrow \phi_{n\pm 1}$. Calculations for non-zero Lorentzian time are performed first in Euclidean time and then analytically continued. This procedure is equivalent to using the Schwinger-Keldysh contour. See \cite{HRTProof} for further discussion of this point. Performing real-time perturbation theory, however, is equivalent to instead using the closed-time (Keldysh) contour and treating the twist operators as well-defined operators in some Lorentzian-signature quantum field theory. See refs. \cite{Weinberg05,Allic16} for a review. While in Lorentzian signature the replica trick is well-defined and twist operators can be identified by their fractional lightcone singularities in correlators, there is no obvious method of defining the twist operators without recourse to complex time.

According to our assumption,
\begin{equation}
\mathcal{I}_\lambda = -\partial_{n=1} \frac{\braket{0|[T(y_-),\Phi_n(u) \bar{\Phi}_{n}(v)]|0}}
{\braket{0|\Phi_n(u) \bar{\Phi}_{n}(v)|0}}.
\end{equation}
We must analytically continue 
\begin{equation}
\mathcal{I}_\lambda =
\left[
\frac{1}{(y_--u)^2}
-\frac{2}{(u-v)(y_--u)}
+\frac{1}{(y_--v)^2}
+\frac{2}{(u-v)(y_--v)}
\right]_{t_y\pm i\epsilon}.
\end{equation}
Using \eqref{TwoPointPrescription},
\begin{equation}
\mathcal{I}_\lambda
=
-\partial_y(\delta(v-y_-)+\delta(u-y_-))
+
\frac{2}{u-v} (\delta(v-y_-)-\delta(u-y_-)).
\numberthis
\label{LorentzianReplicaTrickEE}
\end{equation}
This agrees with the first law result \eqref{FirstLawEE}.

\subsection{AdS$_3$: Hubeny-Rangamani-Takayanagi}

We compute the change to entanglement entropy using the HRT proposal. We first use a bulk diffeomorphism to implement the boundary metric perturbation in analogy to the Euclidean case. Next, we directly compute the geodesic length and find agreement between the two methods and with the CFT result.

In general dimensions, we would need to check that the bulk metric sourced by our metric perturbation at the cutoff surface has a boundary stress tensor expectation value that agrees with the CFT value. However, in pure AdS$_3$, the solution for the bulk metric with a flat boundary CFT metric $g_{\mu \nu}^{(0)} = \eta_{\mu \nu}$ is known exactly. For $\braket{T_{z\bar{z}}}=\braket{T_{\bar{z}z}} =0,$
\begin{equation}
ds^2 = l^2
\left(
L_+ dx_+^2+L_- dx_-^2 - \frac{1}{2}\rho^2 L_+ L_- dx_+ dx_- - 2\frac{1}{\rho^2} dx_+ dx_- + \frac{d\rho^2}{\rho^2} 
\right),
\label{ExactPoincareAdS3Solution}
\end{equation}
where $L_{\pm} \propto \braket{T_{\pm \pm }}$ \cite{Banados98,BalasubramanianK99}. By solving Einstein's equations in the bulk or using the CFT stress tensor two-point function, one can show that perturbations of $g_{\pm \pm }$ are accompanied by a non-zero Weyl anomaly, $\braket{T_{\pm \mp}} \neq 0$ \cite{DHokerKS10}. However, after the perturbation has turned off, $\braket{T_{\pm \mp}} = 0$ and \eqref{ExactPoincareAdS3Solution} applies. It is this regime we are considering.

As in the Euclidean case, we may calculate the change in the HRT surface length by finding the diffeomorphism that reproduces the correct boundary stress tensor expectation values. We will label the boundary lightcone coordinates as $z, \bar{z}$ so the parallels to the Euclidean case are clear.  The small parameter $\epsilon(z,\bar{z})$ is dimensionless and corresponds to the metric perturbation $\delta g_{zz} = \epsilon(z,\bar{z})\delta_{zz}$. We assume $\epsilon$ has compact spacetime support. The diffeomorphism that reproduces $\braket{T(z)} = \partial_z^3 \delta(z-z_s)$ is 
\begin{equation}
\eta \rightarrow \eta+\epsilon l^2 \partial_z \delta(z-z_s) ,
~~~~~
z \rightarrow z - \epsilon  2l \delta(z-z_s),
~~~~~
\bar{z} \rightarrow \bar{z}+ \epsilon l^3 e^{-2 \eta/l} \partial_z^2 \delta(z-z_s).
\label{LorentzianDiffeomorphism}
\end{equation}
As expected, once the source turns off, the bulk metric can be obtained by a diffeomorphism that implements a boundary conformal transformation. The metric becomes 
\begin{align*}
ds^2 =& d\eta^2 
+
e^{2\eta/l} 
dz d\bar{z}
+
\epsilon l^3 \partial_z^3 \delta(z-z_s)
dz dz.
\numberthis
\label{LorentzianMetric}
\end{align*} 
We use the diffeomorphism \eqref{LorentzianDiffeomorphism} to compute the change in entanglement entropy. To first order in the cutoff $\delta$ the extremal surface area changes as
\begin{align*}
S_A &\rightarrow
S_A
+
\frac{c}{3}
\ln\left(1+\epsilon\frac{l}{u-v}
\left(\delta(v-z_s)-\delta(u-z_s)
\right)
\right)
\\
&~~~~~~~~~
-\frac{c}{6}
\left(
\ln(1-\epsilon l \partial_u \delta(u-z_s))
+
\ln(1-\epsilon l \partial_v \delta(v-z_s))
\right).
\numberthis
\end{align*}
Omitting overall factors and using $z_s = y_-$,
\begin{equation}
\mathcal{I}_\lambda
=
-
\partial_{y} (\delta(v-y_-))
+\delta(u-y_-))
+\frac{2}{u-v}
(\delta(v-y_-)-\delta(u-y_-)
)
.
\end{equation}
This agrees with the CFT result \eqref{FirstLawEE}.

A related but distinct computation was performed in \cite{Roberts12}. The method of finding a diffeomorphism to reproduce the AdS$_3$ metric corresponding to a stress tensor perturbation was used to model the time-dependent entanglement entropy of a pulse in a CFT \cite{Roberts12}. In this case the pulse produced a finite expectation value for the boundary stress tensor at all times, while in our case, the stress tensor expectation value turns on at some finite time. The two setups are physically different: in \cite{Roberts12}, the state of the dual CFT being modeled was a mixed state, as the pulse changed the entanglement entropy when its location on the boundary was inside $\mathcal{D}_A$. In contrast, the perturbation we consider is a pure-state perturbation and, as we have seen, does not change the entanglement entropy when the perturbation is within $\mathcal{D}_A$.

We will now reproduce the AdS$_3$ result through directly computing the geodesic length in the background \eqref{LorentzianMetric}. Interestingly, the integral naturally takes the same form as the entanglement first law. To first order in the metric perturbation, the extremal surface does not change. The extremal surface is a geodesic parameterized by $\theta$ as $x = \frac{v+u}{2}-\frac{v-u}{2} \cos \theta, \theta \in [0,\pi]$. With $r = (v-u)/2$, the geodesic is parameterized by $\rho=r \sin \theta$. The new extremal length $\mathcal{L}'$ in terms of the original length $\mathcal{L}$ is 
\begin{equation}
\mathcal{L}'
=\mathcal{L}+\frac{1}{2}
\int d\theta
\frac{\partial_z^3\delta(z-z_s) \left(r \sin\theta \right)^2}
{\sqrt{\frac{1}{\left(r \sin \theta \right)^2}
\left(
(r \cos\theta)^2 + (r \sin \theta)^2
\right)
}}
=
\mathcal{L}+\frac{1}{2}r^2
\int d\theta
(\sin\theta )^3 \partial_z^3\delta(z-z_s).
\end{equation}
Here, $\frac{d}{d z} =  \frac{1}{r \sin \theta}  \frac{d}{d\theta}$. This derivative is singular at the endpoints $\theta = 0, \pi$ as expected. We can rewrite this integral in a more familiar form. Substituting back for $x$ and using
\begin{equation}
dx = -r\sin \theta d\theta,~~~~~~~~~
r^2 \sin^2 \theta = -(x-u)(x-v),
\end{equation}
we can rewrite the integral as an integral over boundary coordinate $x$.
\begin{equation}
\mathcal{I}_\lambda = \int dx \frac{(x-u)(x-v)}{(u-v)} \partial_x^3 \delta(z - z_s). 
\end{equation}
We have demonstrated agreement with the the CFT first law result \eqref{FirstLawBeginning} and therefore \eqref{FirstLawEE}. It would be interesting to extend this to higher order and include matter to obtain higher-order integral expressions for holographic CFT entanglement entropy. By subtracting the contributions from the perturbation to the state, which are known from standard real-time perturbation theory, this procedure would algorithmically calculate the corrections to the expectation value of the modular Hamiltonian for holographic CFTs.

\subsection{Integrating the perturbation and interpretation}

In this section we integrate $\mathcal{I}_\lambda$ against $g$ and discuss the result. As $g$ has compact support, boundary terms arising in integration by parts are zero. We restore the factor of $c$ we had omitted.
\begin{equation}
\delta S_A 
=
c\int dy_+ 
\left[
\partial_v g(v,y_+)+\partial_u g(u,y_+)
+
\frac{2}{u-v} 
\left( 
g(v,y_+) \delta(v-y_-)
-
g(u,y_+) \delta(u-y_-)
\right)
\right].
\label{IntegratedEE}
\end{equation}
We have used the notation $g= g(y_-,y_+)$. The entanglement entropy depends only on $g$ along the lightcones of $u,v$. 

The $\partial g$ term is independent of interval length but zero if $g$ is constant on the lightcone. In the Lorentzian bulk computation, this term arises from changing the cutoff. This interpretation of the result in the CFT is consistent with the $\partial g$ term, as we expect changes in entanglement entropy due to changing the cutoffs at $u,v$ to be additive and independent of the interval length $v-u$. When $g$ changes across the lightcone, that is $\partial_v g(v,y_+) \neq 0$, the relationship between the inner and outer cutoff surfaces changes.

In contrast, the $g$ term changes entanglement entropy even when $g$ is constant across the lightcone. We interpret the $g$ term as arising from correlations of the background state (the vacuum) between different locations on the entangling surface. This interpretation is consistent with the $g$ term, whose contribution decays as $1/(v-u)$ and if $g$ is constant, the change in entanglement entropy along $u$'s lightcone precisely cancels that from $v$'s lightcone. In the bulk calculation, this term comes from transforming the interval length.

Entanglement entropy obeys the causality properties of an operator localized to the entanglement surface. While there are well-known ambiguities in associating entanglement entropy with an observable located at the entangling surface, these ambiguities can arise from gauge invariance \cite{CasiniHR13}. It would be interesting to examine whether these issues arise in computing corrections due to time-dependent perturbations. The causality structure of these corrections depends only on Lorentz-invariant quantities, and so perhaps these causal properties provide a gauge-invariant probe of the physics at the entangling surface.

The result \eqref{IntegratedEE} is valid even when $g(t_y,y)$ is not analytic in time. As previously discussed, Lorentzian-time calculations of entanglement entropy that use the replica trick begin with all operators at zero Lorentzian time and then the result is analytically continued. One may wonder whether this procedure is fundamentally limited, inapplicable when features of the excitation are not analytic in time, for example, discontinuities in the excitation or, in the bulk description, the metric. Perturbatively, however, we see that the computation does proceed by analytic continuation. We do not anticipate that non-analyticity in time poses a fundamental obstruction to implementing the replica trick.

\section{A conjecture: interactions entangle excitations}

The higher-order computations we will perform shortly provide evidence for the conjecture that interactions entangle excitations. In this section we will develop a convenient diagrammatic tool for performing computations in real-time perturbation theory, detail our conjecture, and provide motivation. The content in this section uses real-time perturbation theory \cite{InInSchwinger, InInKeldysh}. For a recent review, see \cite{Weinberg05,Allic16}. In real-time perturbation theory, the time contour in the path integral is never complex.

\subsection{Diagrammatic rules for real-time perturbation theory}
We develop a diagrammatic approach to real-time perturbation theory in order to simplify computations and make basic causality properties manifest. This approach applies to perturbations about a free field theory.

Consider computing corrections to the expectation value of an operator $\mathcal{O}$ due to turning on sources for operators $A,B,C$. We will consider a local operator $\mathcal{O}$ as an example. Spacetime Feynman diagrams describe the various contributions to the integrand at each order, as a function of operator locations $x_A, x_B, x_C, x_{\mathcal{O}}$. In Euclidean AdS/CFT, these diagrams are Witten diagrams. Suppose the contribution we are interested in comes from the commutator $[A,[B,[C,\mathcal{O}]]]]$. The spacetime diagram will be ordered with $t_\mathcal{O}>t_C>t_B>t_A$, which also determines whether operators cross future or past lightcones of other operators when continued to Lorentzian signature. The procedure we give accounts for that sign. The only non-zero contribution to the commutator comes from fully connected contractions. 

\begin{figure}[h]
\begin{center}
\includegraphics[width=0.7 \textwidth]{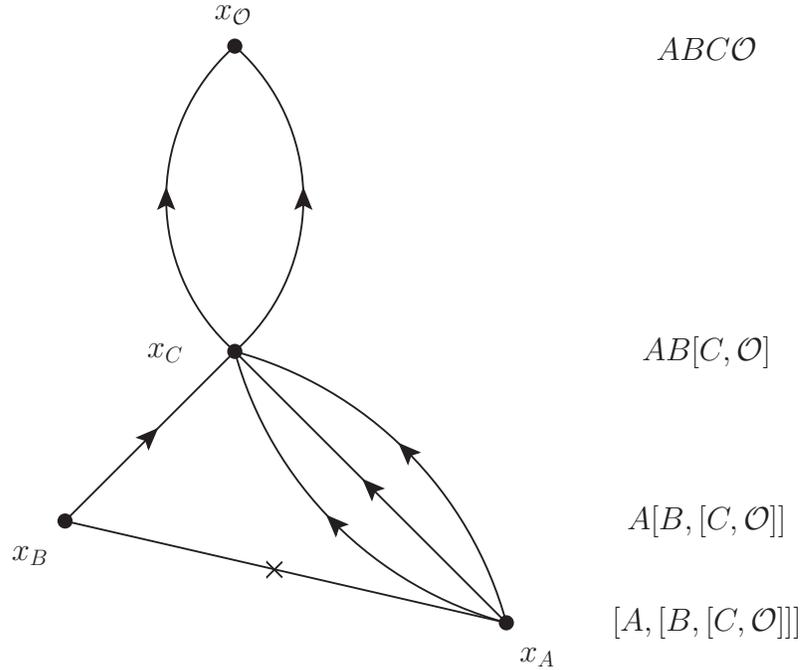}
\end{center}
\caption{\small 
A sample diagram for real-time perturbation theory. We have chosen conventions to make the diagram less visually confusing. When a loop is between null-separated points, we draw only one of the lines null. Some lines therefore may appear spacelike, but it is understood that only the lines labeled with $\times$ are spacelike.
}
\label{SchwingerDiagram}
\end{figure}

In a spacetime diagram, lines that correspond to Wick contractions between operators at spacelike-separated points are labeled separated, as in figure \ref{SchwingerDiagram}. We call these lines spacelike lines for short, and similarly for timelike and null cases. Factor the spacelike contractions $\braket{E}_s$ out of the Euclidean integrand $\braket{E}$, as they will not affect the causal structure of the quantity $\braket{E_c}$ we use to compute the commutator: $\braket{E} = \braket{E}_s \braket{E}_c$. The time-ordering and corresponding commutator can be read off from the spacetime diagram. Begin by continuing $\braket{E}_c$ to Lorentzian time with the following operator ordering:
\begin{equation}
\braket{L}_c^{(1)} = \braket{ABC\mathcal{O}}
\end{equation}
using \eqref{TwoPointPrescription}. Now beginning from $\mathcal{O}$ in the diagram and descending, subtract the continuation corresponding to reversing the $i\epsilon$ associated with each vertex passed. This is nothing more than computing the commutator beginning with $[C, \mathcal{O}]$ and working outwards. Explicitly,
\begin{equation}
\braket{L}_c^{(2)} = \braket{L}_c^{(1)}-\braket{L}_c^{(1)}|_{\epsilon_{\mathcal{O} C} \rightarrow -\epsilon_{\mathcal{O}C}} = \braket{AB[C,\mathcal{O}]}.
\end{equation}
Moving past $B$,
\begin{equation}
\braket{L}_c^{(3)} = \braket{L}_c^{(2)}-\braket{L}_c^{(2)}|_{\epsilon_{\mathcal{O} B},\epsilon_{C B} \rightarrow -\epsilon_{\mathcal{O} B},-\epsilon_{C B}} = \braket{A[B,[C,\mathcal{O}]}.
\end{equation}
Another iteration produces the full Lorentzian commutator
\begin{equation}
\braket{L}_c^{(4)} = [A,[B,[C,\mathcal{O}]]].
\end{equation}
This procedure is illustrated in figure \ref{SchwingerDiagram}.

Other than the rules we have described, the rules for building integrands from diagrams are the standard position-space Feynman rules. In the in-out perturbation theory used for scattering amplitudes, disconnected contractions cancel due to their exponentiation, while here the presence of disconnected contractions makes the whole diagram zero due when the commutators are computed.

Causality properties are manifest in the diagrammatic formulation. Information cannot be transmitted along spacelike lines, but these lines contribute to the perturbative result according to the correlations in the background state. This can be seen in the simple case of the correction to the two-point function given by $[\phi^2(x),[\phi^2(y),\phi(z)\phi(w)]]$. When integrating, spacelike lines can become null or timelike, becoming propagators and carrying information. Non-zero diagrams must be fully-connected to $\mathcal{O}$ once all spacelike lines are cut. From the cutting rule, it follows that every operator insertion must be connected to an operator in its future by at least one null or timelike line. Otherwise, this operator will commute with all operators in its future and the diagram will be zero.

\subsection{The conjecture}

In this section we make a conjecture about entanglement propagation in field theory. We will divide the conjecture into several parts. Consider the Lagrangian of a quantum field theory in $d+1$ dimensions somewhere along its RG flow, written schematically as
\begin{equation}
\mathcal{L} = \mathcal{L}_0 + J(t) \mathcal{O} + \lambda \mathcal{O}_\lambda.
\end{equation}
Assume the initial state $\ket{\Psi}$ is time independent for simplicity.

\begin{figure}[h]
\begin{center}
\includegraphics[width=0.7 \textwidth]{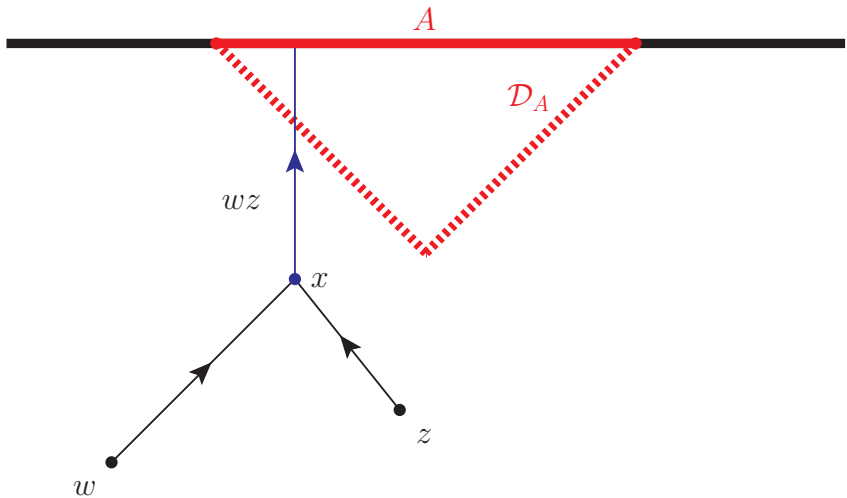}
\end{center}
\caption{\small 
An entanglement diagram for a non-zero $S_{2,1}$ process in $1+1$ dimensions in which propagation occurs only along null rays. Entanglement entropy of a subregion $A$ obeys the causality properties of a non-local operator $\mathcal{O}_{A}$ with support within $\mathcal{D}_A$, and this diagram corresponds to the commutator $[\mathcal{O}(w),[\mathcal{O}(z),[\mathcal{O}_\lambda(x) ,\mathcal{O}_{A}]]]$. The line labeled with $z,w$ is the flow of entanglement of unentangled excitations. If we want to keep track of the flow of all entanglement, we would also include the label $x$. We may also keep track of background state entanglement across spacelike lines.
}
\label{EntanglementDiagram}
\end{figure}

Consider the entanglement entropy $S$ of subregion $A$ at a time $t$ by which the source has turned off: $J(t) = 0$. Suppose $S$ can be expanded in $J, \lambda$:
\begin{equation}
S = \sum_{m,n = 0} S_{m,n} J^m \lambda^n,
\end{equation}
where $S_{0,0}$ is the entanglement entropy in state $\ket{\Psi}$. Suppose that there exists at least one local operator $\mathcal{O}$ such that
\begin{equation}
S_{m,0} = 0.
\end{equation}
Operators $\mathcal{O}$ create unentangled excitations and serve as building blocks for entangled states in that theory. We refer to these operators as unentangled operators, and all others as entangled operators. There generically exists a set of operators $\mathcal{O}_\lambda$ that change the entanglement entropy in the presence of these unentangled excitations:
\begin{equation}
S_{m,n>0} \neq 0.
\end{equation}
When $\mathcal{L}_0$ is a free action, $\mathcal{O}_\lambda$ can sometimes be built from normal-ordered products of $\mathcal{O}$. One can think of operators $\mathcal{O}_\lambda$ as the interactions necessary to entangle unentangled excitations created by $\mathcal{O}$. The first conjecture is that only entangled operators can entangle excitations:
\begin{flalign}
\mathbf{Conjecture~1:}&~~~~~ \text{$\mathcal{O}_\lambda$ is itself an entangled operator}. &&
\label{Conjecture1}
\end{flalign}
$S_{m,n>0} \neq 0$ only when one can draw spacetime ``entanglement diagrams'' to determine for which $n$, interactions $\mathcal{O}_\lambda$, and kinematic regions $S_{m,n>0} \neq 0 $ is allowed. See figure \ref{EntanglementDiagram} for a simple example. The second conjecture is that
\begin{flalign*}
\mathbf{Conjecture~2:}~~~~~ &\text{$S_{m,n>0} \neq 0$ only when an associated entanglement diagram }
\\
&\text{can be drawn of the process.}&&
\numberthis
\label{Conjecture2}
\end{flalign*}
Entanglement diagrams are zero when the same diagram interpreted as a spacetime Feynman diagram would be zero according to the properties explained in Section 5.1.

Even when $\mathcal{L}_0$ is a free Lagrangian, entanglement diagrams generically do not reduce to spacetime Feynman diagrams associated with real-time perturbation theory, whose lines are Wick contractions of local free fields \footnote{In the following section, we will investigate the free fermion using bosonization. In this case, contractions can be taken between the bosons. However, the bosons are not local fields and their relationship to the fermions is inherently non-local as well.}.

One may identify the vertices of entanglement diagrams as follows: for example, in the $S_{3,1}$ correction, the operator $\mathcal{O}_\lambda$ serves as a cubic vertex for operators $\mathcal{O}$ in the associated entanglement diagram when $\braket{\Psi|\mathcal{O} \mathcal{O} \mathcal{O} \mathcal{O}_\lambda|\Psi} \neq 0$. In general,
\begin{flalign*}
\mathbf{Conjecture~3:}~~~~~ &\text{$\mathcal{O}_\lambda$ is an $n$-point vertex for $\mathcal{O}$ in entanglement diagrams when}
\\
&\text{$\braket{\Psi|\mathcal{O}_\lambda(y) \prod_i^n \mathcal{O}(x_i) |\Psi} \neq 0$ at spacelike separations}.&&
\numberthis
\label{VertexCondition}
\end{flalign*}
In large-$N$ field theories, we can choose $\mathcal{O}, \mathcal{O}_\lambda$ to be single-trace primaries whose dimensions are held fixed as $N \rightarrow \infty$; in this case, one consequence\footnote{This follows from large-$N$ factorization of generalized free fields.} of \eqref{VertexCondition} is that $\mathcal{O}_\lambda$ serves as an $m$-point vertex for $m$ odd at order $N^0$. At order $1/N$, $m$ can be even or odd. 

Entanglement diagrams obey the following rules, which are particular to entanglement propagation. At least one timelike or null line must end on the subregion $A$. All $\mathcal{O}_\lambda, \mathcal{O}$ insertions must remain outside the outside $\mathcal{D}_A$\footnote{This fact follows from basic entanglement entropy causality \cite{Headrick14,Allic16}.}. Entanglement diagrams obey an additional cutting rule over for example spacetime Feynman diagrams. Upon cutting all lines that end on $A$, every remaining connected subdiagram must contain at least one operator that is itself entangled. As a corollary, turning on sources for different unentangled operators does not produce entanglement. 

Entanglement diagrams distill two kinds of entanglement: entanglement due to excitations interacting and entanglement due to correlations in the background state. Entanglement cannot propagate along spacelike lines, but correlations in the state $\ket{\Psi}$ will cause excitations to be correlated \footnote{As discussed in Section 5.1, the existence of these two types of correlations is not specific to entanglement entropy, but is a general feature of field theory.}. One can label diagrams according to the propagation of background state entanglement. 

$\mathcal{O}_\lambda$ can entangle excitations that are themselves entangled, but not entangled with each other. This is the more common case, as generic operators are entangled. In this case, $S_{m,0} \neq 0$ but entanglement diagrams still dictate when $S_{m,n} \neq 0$ is allowed and govern the flow of entanglement. We have formulated this section in terms of unentangled excitations, but our statements apply to processes which contain only entangled excitations.

\subsection{Motivation and evidence}

Similar entanglement structure has been found in excited state entanglement entropy computations \cite{HeNTW14,NozakiNT14,Caputa14,HartmanJK15, AsplundBGH115, Allic16}. Working with the free scalar, \cite{NozakiNT14} showed that entanglement entropy in state $e^{i \alpha \phi}\ket{0}$ is equal to that of the vacuum, but in the state $\left(e^{i \alpha \phi}+e^{-i \alpha \phi}\right)\ket{0}$ jumps by $\log(2)$, precisely what is expected from a single entangled pair. The authors put forward a compelling quasi-particle picture, including a discussion of entangled operators. Here, the additional entanglement arose from interactions between the pair of excitations that occurred in preparing the entangled state. In processes with a time-dependent Hamiltonian, the entanglement will arise from interactions for the same reason. We expect the notion of entangled operators creating entangled states to parallel our conjecture for time-dependent Hamiltonians. We have only explored perturbations about the vacuum, but one can perform similar perturbative computations in excited states.

In this work, we consider $\mathcal{L}_0$ for free and holographic  field theories and $\ket{\Psi} = \ket{0}$. Our results in Section 6 will provide evidence for the conjecture in Section 5.2. We consider localized excitations to make the mechanisms physically transparent. Without a general method to compute field theory entanglement entropy, it is unclear how to prove the statements in Section 5.2 that we have not already shown to follow from basic properties of entanglement and causality. While entanglement entropy is not itself a physical observable, it is determined by $\rho_A$, and all observable properties of $\rho_A$ are fixed according to the expectation values of operators localized to $\mathcal{D}_A$, which themselves change according to standard real-time perturbation theory.

\section{Higher order perturbation theory: the free fermion}

The computations in this section demonstrate the mechanisms that we conjecture in Section 5.2. Even in the absence of excitations, deforming a CFT by some operator will change its vacuum entanglement entropy. This can be seen from the Euclidean perturbation theory and has been well studied. How this happens is clear: deforming the CFT changes the vacuum state. Here, we will perform computations that reveal a different mechanism: separately from changing the vacuum state of the theory, interactions change entanglement entropy by entangling excitations.

We calculate higher-order corrections to entanglement entropy in the free $1+1$ dimensional fermion using the replica trick. The twist operators for the free fermion are known explicitly and so the result can be computed exactly, with causality properties manifest at every step. We compute several entanglement diagrams built from the following chiral operators:
\begin{equation}
J = \sum_k \partial \phi_k,~~~~~ T = \sum_k (\partial \phi_k)^2,~~~~~ W = \sum_k (\partial \phi_k)^3,
\end{equation}
with $k = -1/2(n-1),-1/2(n-1)+1,\ldots,1/2(n-1)$. These operators are the spin 1, 2, and 3 currents respectively, which are built from bilinears of fermion fields, schematically $:\psi \partial^m \psi^*:$ with $m=0,1,2$ \cite{Pope1991,DattaDFK14}. 

\subsection{Warmup: metric perturbation}

To understand basic features of fermionic calculations of entanglement entropy, we warm up by reproducing the first-order change in entanglement entropy due to a metric perturbation \eqref{LorentzianReplicaTrickEE}. Using \eqref{NormalOrderedTwists},
\begin{equation}
\braket{0|T(z) \Phi_n(u) \bar{\Phi}_n(v)|0}
=
\sum_k \frac{-k^2}{n^2}
\braket{0|:(\partial \phi_k)^2(z):
:
\left(
\phi_k(u)
-
\phi_k(v)
\right)^2:
|0}.
\end{equation}
The singularity structure of \eqref{LorentzianReplicaTrickEE} is apparent even before acting with $-\partial_{n=1}$ to obtain entanglement entropy.
\begin{equation}
\braket{0|T(z) \Phi(u) \bar{\Phi}_n(v)|0}
=
\sum_k \frac{-k^2}{n^2}
\left(
\frac{1}{(z-u)^2}
+
\frac{1}{(z-v)^2}
+
\frac{2}{(u-v)(z-v)}
-
\frac{2}{(u-v)(z-u)}
\right).
\label{FermionStressTensorPerturbation}
\end{equation}
The remaining steps lead to \eqref{LorentzianReplicaTrickEE}.

\subsection{$J$ creates unentangled excitations}

We will compute entanglement entropy due to perturbations in $J$ with a time-dependent source. Without loss of generality, we will suppose $v$ is spacelike-separated from all sources to simplify the expressions unless specified otherwise. At first order in $J(z)$, entanglement entropy does not change. The Euclidean integrand is
\begin{equation}
\mathcal{I}_{J}^E(z) = \sum_k \frac{k}{n}\braket{0|\partial \phi_k (z)\phi_k(u)|0}
=
\sum_k \frac{k}{n} \frac{1}{z-u}.
\end{equation}
Performing the sum over $k$, we see $\mathcal{I}_{J}=0$, as expected for a primary operator. The second-order correction comes from 
\begin{equation}
\mathcal{I}_{J^2}^E(w,z) 
=
\sum_{k_1, k_2,k_3,k_4} \braket{0|\partial \phi_{k_1}(w)\partial \phi_{k_2}(z):\frac{k_3}{n}\phi_{k_3}(u) \frac{k_4}{n}\phi_{k_4}(u):|0}.
\end{equation}
Performing the sum over $k_3,k_4$ gives zero in any connected correlator above. All higher-order corrections will be zero for the same reason, which is that $J$ is linear in $\partial \phi$. $J$ therefore creates an unentangled excitation as defined in Section 5.2. $J$ is indeed a non-trivial excitation, as it does change correlators of fermions on the plane.

\subsection{Adding an interaction $W$ entangles $J$ excitations}

We introduce a cubic interaction $W$ and observe how this entangles two unentangled $J$ excitations. To $\mathcal{O}(J^2 W)$,
\begin{equation}
\mathcal{I}_{J^2 W}^E(w,z,x) 
=
\sum_{k_1, k_2,k_3} \braket{0|\partial \phi_{k_1}(w)\partial \phi_{k_2}(z):(\partial \phi_{k_3})^3(x)::e^{\sum_{k_4} i \frac{k_4}{n}\phi_{k_4}(u)}:|0}.
\end{equation}
All connected contractions are zero because they involve a sum over an odd power of $k_i$. Non-zero correlators require an even number of $\phi$ operators within the correlator with twist operators. At $\mathcal{O}(J^2 W^2)$,
\begin{align*}
\mathcal{I}_{J^2 W^2}^E(w,z,x,y) 
&=
\sum_{k_1, k_2,k_3,k_4} \bra{0}\partial \phi_{k_1}(w)\partial \phi_{k_2}(z)
\\
&~~~~~~~~~ \times
:(\partial \phi_{k_3})^3(x)::(\partial \phi_{k_4})^3(y)::e^{\sum_{k_5} i \frac{k_5}{n}\phi_{k_5}(u)}:\ket{0}.
\numberthis
\end{align*}
Not all contractions contribute to entanglement entropy. All terms with a single contraction between $\phi(w)$ or $\phi(z)$ with $\phi(u)$ are zero upon summing over $k_i$. This is true to all orders in $J,W$, consistent with the cutting rule for entanglement diagrams, as $J$ represents an unentangled excitation. This conjectured cutting rule is non-trivial and specific to entanglement entropy: in computing corrections to generic correlators instead of entanglement entropy, these contractions would not be zero. If all lines ending on $A$ are cut and the diagram $D$ factorizes into $D_{ent} D_{unent}$, which is the product of a diagram that changes the entanglement entropy on its own and a diagram containing only unentangled excitations, then the associated change in entanglement entropy occurs at a different order in perturbation theory than the original diagram.

\begin{figure}[h]
\begin{center}
\includegraphics[width=0.7 \textwidth]{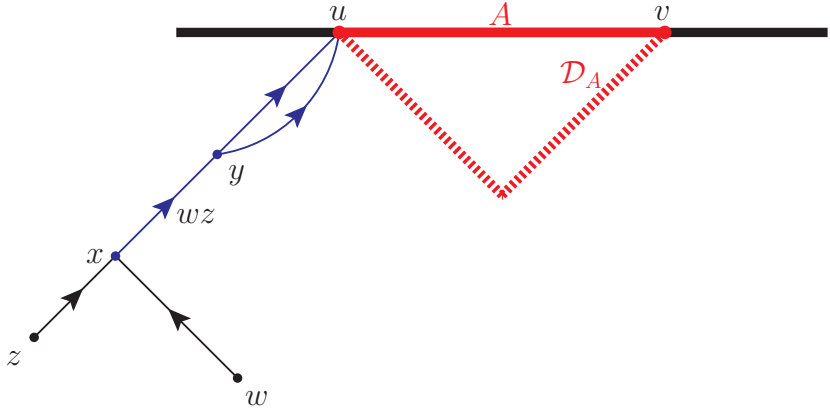}
\end{center}
\caption{\small 
A non-zero diagram contributing to the $\mathcal{O}(J^2 W^2)$ process. We have drawn one of the incoming lines as left-moving for clarity.
}
\label{FermionEntanglementDiagramNonzeroQuantumEntanglement}
\end{figure}

The diagram in figure \ref{FermionEntanglementDiagramNonzeroQuantumEntanglement} with the $x,y$ contractions exchanged has no branch cuts in $u-y$, and so will be zero once we compute the corresponding commutators. As explained in Section 5.1, this is the real-time perturbation theory rule that every vertex must have at least one future-directed line that is not spacelike.

\begin{figure}[h]
\begin{center}
\includegraphics[width=0.7 \textwidth]{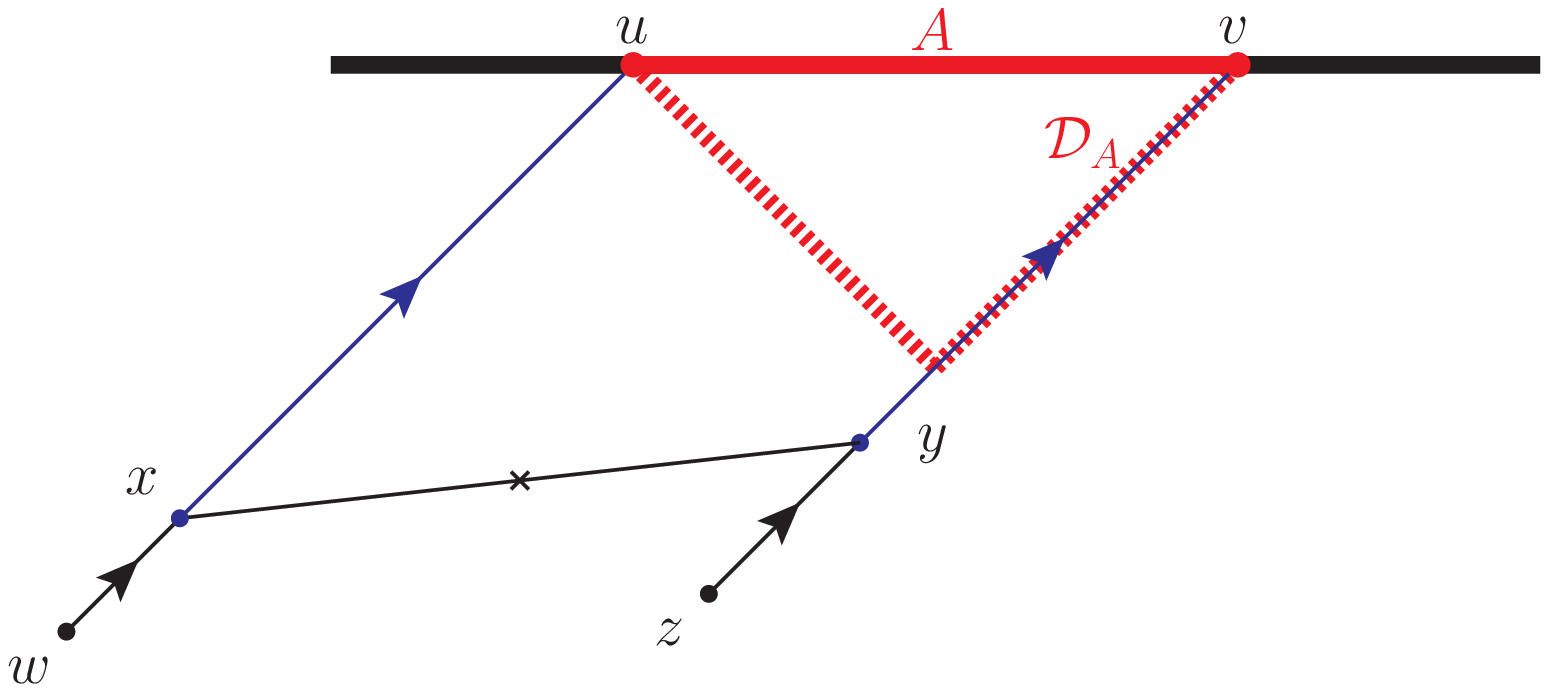}
\end{center}
\caption{\small 
A diagram contributing to the $\mathcal{O}(J^2 W^2)$ process. This diagram corresponds to a change in entanglement entropy due to background state correlations. A similar diagram contributes when $x,y = u$, in which case the entanglement is not due to background state correlations.
}
\label{FermionEntanglementDiagramNonzeroClassicalEntanglement}
\end{figure}

There are two allowed diagrams, and which one is non-zero depends on the location of $x,y$. The diagram in figure \ref{FermionEntanglementDiagramNonzeroQuantumEntanglement} corresponds exclusively to the causal entanglement of excitations. In contrast, when for example $u=x$ and $v=y$, the diagram in figure \ref{FermionEntanglementDiagramNonzeroClassicalEntanglement} corresponds exclusively to background state entanglement.

We compute the diagram in figure \ref{FermionEntanglementDiagramNonzeroQuantumEntanglement}. The Euclidean signature integrand is
\begin{equation}
\mathcal{I}_{J^2 W^2}^E(w,z,x,y) = \frac{1}{(w-x)^2(z-x)^2(x-y)^2(y-u)^2}.
\end{equation}
Using the procedure in Section 5.1, it is straightforward to compute the commutator, although we will shortly use real-time Feynman rules to simplify the process further. Using the top sign in the top line of \eqref{TwoPointPrescription},
\begin{align*}
\mathcal{I}_{J^2 W^2}^{(1)}(w,z,x,y) &= 
\left(\frac{1}{(w-x)^2}+i \pi \partial_{w-x} \delta (w-x)\right) 
\left(\frac{1}{(z-x)^2}+i \pi  \partial_{z-x}\delta (z-x)\right)
\\
&~~~~~\times
\left(\frac{1}{(x-y)^2}+i \pi \partial_{x-y}\ \delta (x-y)\right)
\left(\frac{1}{(y-u)^2}+i \pi \partial_{y-u} \delta (y-u)\right)  .
\numberthis
\end{align*}
As in Section 5.1, we begin with the operator at $u$ descend to compute the commutator.
\begin{equation}
\mathcal{I}_{J^2 W^2}(w,z,x,y) = \partial_{w-x}\delta (w-x) \partial_{z-x}\delta (z-x)
 \partial_{x-y}\delta (x-y) \partial_{y-u}\delta (y-u).
\end{equation}
We see that the $W$ operators entangle the unentangled $J$ excitations and entanglement travels according to the associated entanglement diagram. It is straightforward to integrate this result against sources for the operators, as in \eqref{IntegratedEE}.
\begin{equation}
\delta S_{J^2 W^2} = \int d^2w d^2z d^2x d^2y \partial_w \mathcal{J}_J(w,\bar{w}) \partial_{z} \mathcal{J}_J (z,\bar{z}) \partial_x \mathcal{J}_W (x,\bar{x}) \partial_y \mathcal{J}_W (y,\bar{y})\bigg|_{w=z=x=y=u},
\end{equation}
where $\mathcal{J}_J, \mathcal{J}_W $ are the source functions for operators $J, W$.

Because we work in free field perturbation theory, real-time position space Feynman rules apply, and they simplify calculations further. We will need the Feynman rules for operators $\phi, \partial \phi$. When spacetime points $a,b$ are causally connected, $\phi(a) \partial \phi(b)$ contributes a $\delta(a-b)$ and $\partial \phi(a) \partial \phi(b)$ contributes $\partial_{a-b} \delta(a-b)$. When $a,b$ are spacelike-separated, $\phi(a) \partial \phi(b)$ contributes $1/(a-b)$ and $\partial \phi(a) \partial \phi(b)$ contributes $1/(a-b)^2$. We have omitted the overall numerical factors in these expressions.

Using the above Feynman rules, we can now easily compute the diagram in figure \ref{FermionEntanglementDiagramNonzeroClassicalEntanglement} in the kinematic regime $w=u,z=u$. We find the contribution
\begin{equation}
\mathcal{I}_{J^2 W^2}(w,z,x,y) = \frac{1}{(y-x)^2}
\partial_{w}\delta (w-x)\delta (x-u)  
\partial_{z}\delta (z-y) \delta (y-v). 
\end{equation}
We see that the background state entanglement decays in $y-x$. That this diagram decays in $y-x$ is expected from the spatial decay of vacuum correlations. In the regime $w=z=u$,
\begin{equation}
\mathcal{I}_{J^2 W^2}(w,z,x,y) = 
\delta (y-u) \delta(x-u)
\partial_{y} \delta(y-x)
\partial_{z}\delta (z-y) 
\partial_{w}\delta (w-x). 
\end{equation}
At $\mathcal{O}(J^2 W^2)$ performing the integration reveals similar features to the metric perturbation result \eqref{IntegratedEE}, namely that the sources often must change across the lightcones of $u,v$ in order to change entanglement entropy. This is a distinctly time-dependent feature of entanglement entropy and its interpretation is similar to that of \eqref{IntegratedEE}. Diagrams in other kinematic regions are straightforward to obtain from the results we have given, and higher-order integrands are similarly easy to construct using the tools we have provided.

As all the operators we have used are fermion bilinears, the standard Wick contraction between fermions would not have produced a diagram with a cubic vertex. However, the condition we gave in \eqref{VertexCondition} identifies $W$ as a cubic vertex. The bosonic field $\phi$ and entanglement diagrams naturally describe entanglement propagation similarly to how the fermionic field $\psi$ and spacetime Feynman diagrams naturally describe time-dependent corrections to local observables. It is straightforward to show that $W$ creates entangled excitations, and so the conjectures \eqref{Conjecture1}, \eqref{Conjecture2}, and \eqref{VertexCondition} hold for operators $J,W$.

\section{Discussion}

In this work, we have investigated entanglement entropy in conformal field theory with a general time-dependent Hamiltonian. We have extended previous first-order studies by computing higher order corrections. Past first order, we found evidence of a universal entanglement structure in perturbation theory. We have conjectured that interactions entangle unentangled excitations. This conjecture has a practical use: it provides a prescription for identifying the building blocks of entanglement according to a microscopic description of how these building blocks interact to generate entanglement. Using the free fermion as an illustrative example, we identified the spin-1 current $J$ as an unentangled excitation, included the spin-3 current $W$ as an entangled interaction and found that, when the interaction turns on where the two $J$ excitations collide, the interaction entangles the excitations. We computed the corresponding $J^2 W^2$ processes. Having identified unentangled excitations $J$ and an entangled cubic vertex $W$, we show that the free fermion is a simple, tractable arena in which to investigate details of entanglement propagation. We provided a diagrammatic approach to real-time perturbation theory and found this approach makes causality properties of the correlators manifest, as well as streamlines their computation. 

We have conjectured that the flow of entanglement entropy is governed by ``entanglement diagrams''. Entanglement propagates only when there is a non-trivial entanglement diagram associated with the process. These diagrams are motivated by the accessory spacetime Feynman diagrams of real-time perturbation theory, but they do not correspond to Feynman diagrams of local operators. Entanglement diagrams obey non-trivial rules specific to entanglement entropy. In the bosonized free fermion, entanglement diagrams are spacetime Feynman diagrams but for the boson $\phi$ rather than the fermion $\psi$. The procedure we propose for identifying vertices in entanglement diagrams identifies $W$ as a cubic vertex independently of the bosonization procedure that makes this fact manifest. More generally, we expect that further study of entanglement diagrams and the entanglement of excitations by entangled interactions may uncover natural variables for entanglement propagation in field theory.

\subsection{Future directions}

We detail several exciting directions of further study, some of which we hope to report on in the future.

Causality may place stringent constraints on the $1+1$ dimensional modular Hamiltonian via a bootstrap approach, order by order in perturbation theory. In Section 4 we found that entanglement causality places a non-trivial constraint on the integrand of the modular Hamiltonian if the integrand is function of local operators. Using the replica trick and Ward identities, one can compute the entanglement entropy due to metric perturbations to arbitrary order. Constraints on the modular Hamiltonian at a given order feed into its form at the next order. The ansatz can be checked for consistency against the exact expression, obtainable using standard perturbation theory \cite{FaulknerHHPRR17}. At each order, the modular Hamiltonian must be consistent with entanglement causality due to perturbation by primary operators as well.

While local, explicit twist-operators are special to $1+1$ dimensional fermions, replica trick computations in the free scalar and fermions are in principle straightforward in general dimensions for a single interval \cite{NozakiNT14}. One can clarify the relationship between entangled operators for excited states and Hamiltonian perturbations, as there is a correspondence between local operator excitations and Hamiltonian perturbations \cite{Allic16}. For example, vertex operators have a simple interpretation as building blocks of EPR states \cite{NozakiNT14}. What is the entanglement propagation structure of vertex operators as Hamiltonian perturbations? The entangling interactions $\mathcal{O}_\lambda$ affect entanglement propagation. A concrete question to answer is: how does entanglement velocity depend on the choice of $\mathcal{O}_\lambda$? We have only addressed entanglement entropy, but one may investigate similar questions for other information theoretic measures.

Time-dependent Hamiltonian perturbations can serve as another probe of HRT. Specifically, approximate expressions for the conformal blocks dominating the $\braket{\mathcal{O} \mathcal{O} \Phi_n \bar{\Phi}_n}$ correlator are known in the large $c$ limit for holographic CFTs \cite{AsplundBGH215, FitzpatrickKW14,FitzpatrickKW15}. This correlator gives the entanglement entropy to second order in the Hamiltonian perturbation $J(t) \mathcal{O}$, and one can compare the result to the bulk HRT calculation. Keeping $\Delta_{\mathcal{O}}/ c$ fixed in the large-$c$ limit corresponds to a non-perturbative bulk computation. In \cite{AnousHRS16} one part of this CFT calculation was performed. $\Delta_{\mathcal{O}}(n) = n \Delta_{O}$ was used, while the full second order correction involves the operator $\sum_k^n J^k \mathcal{O}_{k}$, where $\Delta_{O_k} = k \Delta_{\mathcal{O}}$. On the other hand, holding $\Delta_{\mathcal{O}}$ fixed in the large-$c$ limit corresponds to standard bulk semiclassical perturbation theory.

While we have so far discussed entanglement in flat spacetimes, it is well known that the structure of entanglement in AdS plays an important role in the AdS/CFT correspondence. In particular, one may investigate the entanglement spreading that we have described in flat space in AdS, and address its dual CFT interpretation. What is the CFT dual of bulk entanglement diagrams like figure \ref{EntanglementDiagramIntroduction}? Specifically, tree-level Witten diagrams can be expressed as linear combinations of conformal blocks, see for example \cite{GeodesicWittenDiagrams}, and so we expect that bulk tree-level entanglement diagrams have a dual CFT description at the corresponding order in $1/N$. How do the UV data of the theory - the interactions - affect entanglement structure on a covariant measure of entanglement and entropy, namely light sheets \cite{CovariantEntropyBound,BoussoCFM14}?

One can study loop-level effects in entanglement entropy. The loop-level integrals are challenging, but these may be addressed using modern machinery developed to calculate scattering amplitudes \cite{ElvangH13,Henn14,Smirnov05}. Integrating over the external sources is a task particular to real-time perturbation theory, but computing entanglement entropy order by order in interactions $\mathcal{O}_\lambda$ amounts to computing loop entanglement diagrams, and the integrals are analytic continuations of those that appear in loop-level scattering processes. Using the free scalar and fermion, and the perturbation theory tools we have presented here, one can investigate how UV behavior affects entanglement entropy.

\section{Acknowledgments}
It is a pleasure to thank Per Kraus for many constructive discussions and a careful reading of a draft of this manuscript. We thank Eliot Hijano and Christoph Uhlemann for comments on the manuscript.

\bibliographystyle{ssg}
\bibliography{refs}

\end{document}